# Non-diffusion transport in decoherent non-Hermitian quasicrystals

Yudong Ren[1,2,3], Rui Zhao[1,2,3], Kangpeng Ye[1,2,3], Lu Zhang[1,2,3], Hongsheng Chen[1,2,3,*], Haoran Xue[4,5,*], Yihao Yang[1,2,3,*]

[1]State Key Laboratory of Extreme Photonics and Instrumentation, ZJU-Hangzhou Global Scientific and Technological Innovation Center, Zhejiang University, Hangzhou 310027, China.
[2]The Electromagnetics Academy at Zhejiang University, College of Information Science and Electronic Engineering, Zhejiang University, Hangzhou 310027, China.
[3]Zhejiang Key Lab. of Intelligent Electromagnetic Control and Advanced Electronic Integration, Jinhua Institute of Zhejiang University, Zhejiang University, Jinhua 321099, China
[4]Department of Physics, The Chinese University of Hong Kong, Shatin, Hong Kong SAR, China.
[5]State Key Laboratory of Quantum Information Technologies and Materials, The Chinese University of Hong Kong, Shatin, Hong Kong SAR, China

* Corresponding author. Email: yangyihao@zju.edu.cn (Y. Y.); haoranxue@cuhk.edu.hk (H. X.); and hansomchen@zju.edu.cn (H. C.).

**Abstract**

Disorder and coherence jointly govern wave transport in complex media. In Hermitian systems, a long-established paradigm since Anderson's work holds that disorder-induced localization relies on phase-coherent interference, and that the loss of coherence inevitably suppresses localization and restores featureless diffusive transport at long times. Whether this intuition remains valid in non-Hermitian systems, where transport can be governed by dissipation rather than interference, has remained largely open. Here we theoretically and experimentally demonstrate that this paradigm fundamentally breaks down in decoherent non-Hermitian quasicrystals. Using a programmable photonic lattice with independently engineered dissipation and fully programmable dephasing, we access regimes spanning from fully coherent to fully incoherent dynamics. While decoherence washes out localization and enforces structureless diffusion in Hermitian lattices, we find that decoherent non-Hermitian quasicrystals retain nontrivial, non-diffusive transport structures even in the incoherent limit. These include dissipation-induced localization, diffusion-localization transitions, and decoherence-induced mobility edges, phenomena with no counterparts in Hermitian disordered systems. We develop a unified theoretical framework that captures the ensemble-averaged dynamics across the entire coherence landscape, continuously connecting coherent and incoherent regimes, and reveals how dissipation and decoherence cooperate to shape transport. Our results establish decoherent non-Hermitian lattices as a distinct class of transport systems, in which dissipation and incoherence generate structured, non-diffusive phases, beyond the conventional Anderson picture.

## MAIN TEXT

### Introduction

Disorder and coherence play central roles in determining the transport properties of wave systems. Before Anderson's seminal work[1], disorder in electronic systems was primarily viewed as a source of enhanced scattering and resistance, while diffusion was still assumed to govern transport in the long-time limit (Drude model[2] and Sommerfeld theory[3]). Anderson localization fundamentally overturned this intuition by demonstrating that, in coherent wave systems, disorder can halt diffusion completely and lead to exponential wavefunction localization. Originally discovered for electrons, this phenomenon was extended to photons[4–8], sound[9], and matter waves[10–12], establishing phase-coherent interference as the universal mechanism behind disorder-induced localization and enabling devices such as coherent-feedback random lasers[13]. Crucially, this mechanism depends on preserved phase relationships: interference among multiple scattering paths sustains localization, while temporal fluctuations, environmental noise, or dynamical disorder progressively destroy the phase correlations[14–20], and even weak dephasing is sufficient to suppress Anderson localization and restore diffusive transport at long times. This behavior has been consistently observed across electronic, photonic, and other wave platforms[17,19,21,22], leading to a widely accepted paradigm in disordered wave physics: in Hermitian systems, the loss of coherence inevitably drives transport toward diffusion as the universal asymptotic behavior.

Non-Hermitian physics[23] introduces a fundamentally different route to controlling transport, which does not rely solely on phase-coherent interference[24–26]. A first striking example is the non-Hermitian skin effect[27], which features the breakdown of conventional bulk-boundary correspondence and is linked to a nontrivial point-gap spectral topology[28]. Its hallmark is a persistent directional current in the bulk dynamics[29], demonstrating how non-reciprocity alone can dramatically redirect and localize waves, a control mechanism distinct from interference-based localization[23,30]. Moving to disordered settings, another key example is that non-Hermitian disorder can localize eigenstates through a mechanism distinct from the Hermitian Anderson picture[31]. Even when all eigenstates are exponentially localized by disorder, wave propagation can still be dictated not by their spatial profiles, but by competition among them[32], where the universal long-time transport is governed by the tails of the imaginary-part density of states[33]. These phenomena have no direct Hermitian counterparts and are not rooted solely in interference. These examples suggest that in non-Hermitian systems, transport can be governed by mechanisms beyond coherent interference, raising the possibility of a fundamentally different interplay between disorder and decoherence. This naturally leads to a fundamental and largely unexplored question: in non-Hermitian decoherent systems, where both disorder and dissipation are at play, does the loss of phase coherence still restore diffusion, or can it instead cooperate with non-Hermitian effects to give rise to unexpected, non-diffusive phases (Fig. 1a)?

Here we address this question by constructing a programmable photonic lattice that allows independent control over non-Hermitian dissipation, coherent coupling, and spatiotemporally random phases. This platform provides access to the full coherence landscape spanning from fully coherent to completely incoherent limits, thereby allowing us to systematically probe how dephasing interacts with non-Hermitian order. With this platform, we experimentally observe that the decoherent non-Hermitian quasiperiodic lattice behaves in a fundamentally different manner from its Hermitian counterpart. These behaviors include dissipation-induced localization, diffusion-localization transitions, and decoherence-induced mobility edges, phenomena without counterparts in Hermitian systems. We also develop a unified theoretical framework that captures the exact ensemble-averaged dynamics from coherent evolution to completely incoherent propagation, providing an analytical foundation for these observations.

## Results

### A non-Hermitian quasicrystal with programmable dephasing

We realize a programmable photonic lattice using two coupled optical fiber loops with unequal lengths (Fig. 1b, also see Extended Data Fig. 1 for details), which offers an excellent testbed to simulate intriguing effects, such as parity-time symmetry[34], non-Hermitian skin effect[35], negative optical temperature[36], and gain/loss-induced Dirac mass[37]. When optical pulses propagate through the two loops, they arrive at the coupler (i.e., a variable beamsplitter in our setup) at different times. The delay between the pulses propagating within the two loops is utilized to create synthetic dimensions[38] (see Methods). The initial pulse repeatedly splits at the variable beamsplitter and, after several round trips, multipath interference between the emerging sub-pulses takes place. This pulse evolution realizes a photonic quantum walk in the synthetic dimension. Thus, this system can be mathematically mapped into a double-discrete 1+1D mesh lattice with a longitude (time) dimension and a transverse (space) dimension, as shown in Fig. 1c.

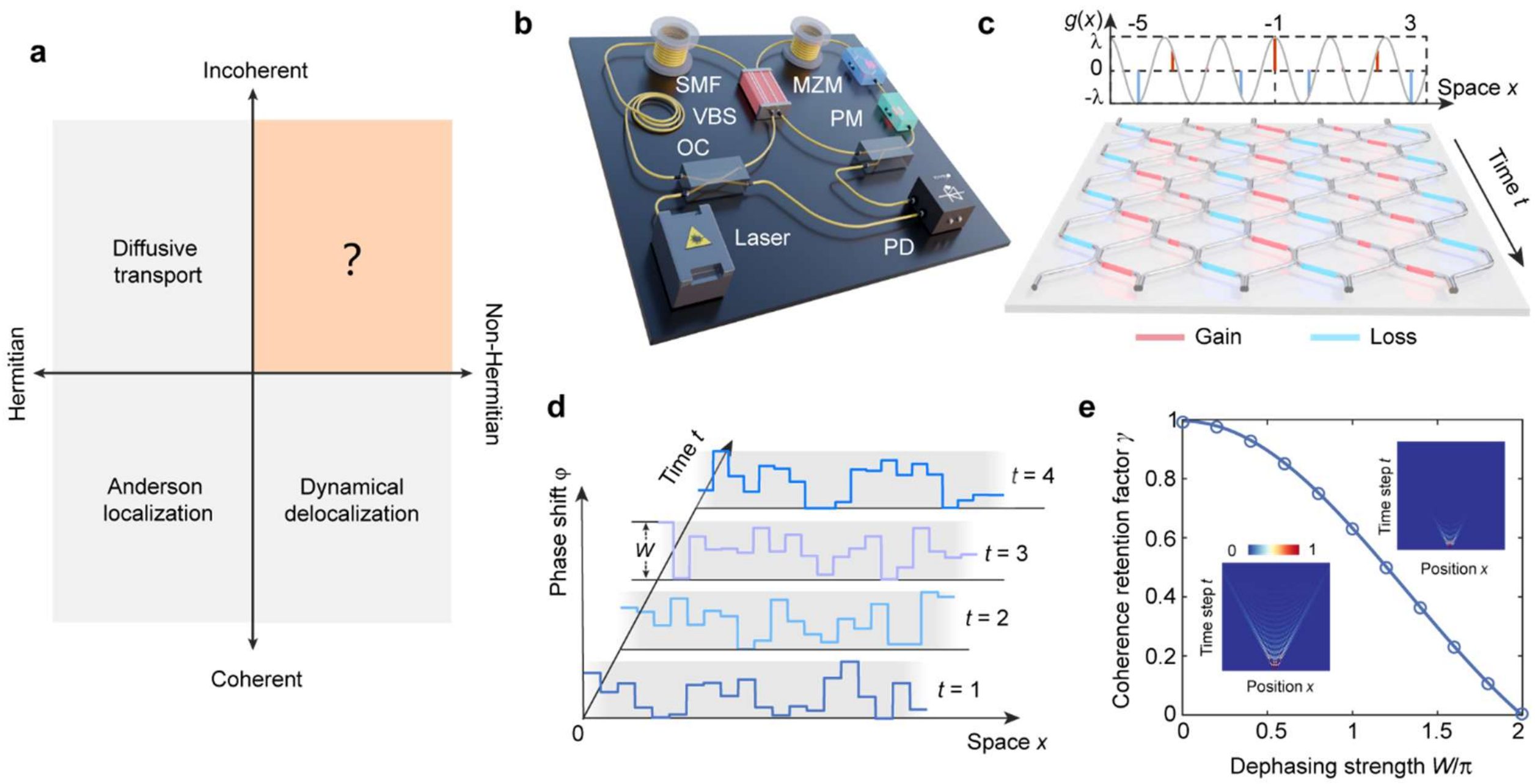

**Fig. 1. Concept and experimental platform.** (**a**) Diagram illustrating regimes of wave transport in disordered systems, mapped onto the coherence-Hermiticity plane. Different combinations give rise to different hallmark phenomena: diffusion in incoherent Hermitian systems, Anderson localization in coherent Hermitian systems, and dynamical delocalization in coherent non-Hermitian systems. The fourth quadrant (non-Hermitian and incoherent) remains largely uncharted. (**b**) Schematic diagram of the experimental setup. The system consists of two fiber loops with different lengths. PM: phase modulator; MZM: Mach-Zehnder modulator; OC: optical coupler; VBS: variable beamsplitter; SMF: single-mode fiber; PD: photodetector; (**c**) Photonic mesh lattice with non-Hermitian quasicrystalline order. Quasiperiodic dissipation is implemented by modulating the intensity of the left-propagating pulse at each site $x$. (**d**) Implementation of programmable dephasing. The phase shift $\phi(x,t)$ is randomly modulated in both time step $t$ (vertical) and lattice site $x$ (horizontal), introducing controlled dephasing across the system. (**e**) Coherence retention factor $\gamma$ as a function of dephasing strength $W$. $\gamma$ quantifies the degree of dephasing, bridging the fully coherent ($W = 0$, $\gamma = 1$) and fully incoherent ($W = 2\pi$, $\gamma = 0$) limits. Insets show simulated mean-field patterns at $W = 0.2\pi$ (left) and $W = 0.4\pi$ (right).

Crucially, the platform allows independent and real-time control of both the Hermitian and non-Hermitian parts of the lattice potential. A Mach-Zehnder modulator (MZM) in one loop imposes a site-dependent, incommensurate gain-and-loss profile $g(x) = \lambda\cos[2\pi\alpha(x+1)]$, where α is an irrational number that sets the quasiperiodicity. Here, λ quantifies the amplitude of the non-Hermitian modulation, and this term realizes the imaginary quasiperiodic potential of a non-Hermitian Aubry–André model (Fig. 1c). Simultaneously, a phase modulator (PM) introduces dynamic phase shifts that fluctuate randomly in both the time step $t$ and the lattice site $x$ (Fig. 1d). These shifts deliberately destroy the phase coherence between different propagation paths, thereby realizing controlled dephasing (or decoherence) in the photonic lattice[17,39,40]. Together, the MZM and PM provide independent, real-time control over the imaginary quasiperiodic potential and the strength of dephasing, allowing us to continuously explore the interplay between non-Hermitian order and decoherence. The discrete-time evolution of the pulse amplitudes in the two channels is governed by

$$\begin{aligned} u(x,t+1) &= [\cos(\theta)u(x+1,t) + i\sin(\theta)v(x-1,t)]e^{g(x)+i\phi(x,t)} \\ v(x,t+1) &= i\sin(\theta)u(x-1,t) + \cos(\theta)v(x+1,t), \end{aligned} \tag{1}$$

where $u(x,t)$ and $v(x,t)$ denote the field amplitudes on the left-moving (short-loop) and right-moving (long-loop) channels, at position $x$ and time step $t$, respectively. $\theta$ is the splitting ratio of the variable beamsplitter, which also corresponds to the coupling strength in the effective lattice model. The function $\phi(x,t)$ represents the dynamically randomized phase that induces dephasing. The function $g(x)$ provides the quasiperiodic imaginary potential, which together with the random phases $\phi(x,t)$ establishes a non-Hermitian quasicrystal subject to fully programmable dephasing.

**A unified theoretical framework for decoherent lattice dynamics**

To quantitatively capture the ensemble-averaged dynamics and the corresponding eigenspectrum under arbitrary dephasing strength, we develop a general theoretical framework for the decoherent lattice considered here. In our model, dephasing is introduced through random phases $\phi(x,t)$ that are uncorrelated in both time step $t$ and lattice site $x$. These phases are uniformly distributed in $[-W/2, W/2]$, where the parameter $W \in [0, 2\pi]$ quantifies the dephasing strength. We characterize the decoherent dynamics through a hierarchy of correlation functions. The central quantity is the two-point correlation tensor $\mathcal{F}_n^{\alpha\beta}(x,t) = \langle \alpha(x+n,t)\beta^*(x-n,t)\rangle$, where the $n$ denotes the correlation order (spatial separation), and $\alpha$, $\beta \in \{u, v\}$ label the internal degrees of freedom (short- and long-loop channels). Here $\langle\cdot\rangle$ denotes the ensemble average. The zeroth-order components ($n = 0$) correspond to the on-site densities $\langle|u(x,t)|^2\rangle$ and $\langle|v(x,t)|^2\rangle$, while higher-order terms encode phase coherence over increasingly distant sites. The evolution of these correlators is governed by a compact recursive equation (see Supplementary Note 1),

$$\mathcal{F}_n^{\alpha\beta}(t+1,x) = \Phi_{\alpha\beta}(x,n)\Gamma_{\alpha\beta}\, R^{\alpha\beta}_{\alpha'\beta'}\, \mathcal{F}^{\alpha'\beta'}_{n+\Delta n(\alpha\beta)}(t, x+\Delta x(\alpha\beta)). \tag{2}$$

Here, $R(\theta)$ is the coupling tensor determined by the beamsplitter ratio $\theta$, which describes coherent scattering between channels; $\Gamma(W)$ is the coherence tensor arising from dephasing. In the symmetric case where noise is applied to both $u$ and $v$ components, $\Gamma(W) = \gamma^2(W)$, where $\gamma(W) = 2\sin(W/2)/W$ is the coherence retention factor, quantifying how much phase coherence survives after averaging over dephasing configurations. In our experiment and the subsequent analysis, we consider the asymmetric case with noise applied only to the $u$ component. Here, the tensor $\Gamma(W)$ takes a slightly complicated form (see Supplementary Note 1), yet coherence decay remains characterized by the same factor $\gamma$. Physically, $\gamma$ governs the decay of coherence: the mean fields $\langle u(x,t)\rangle$ and $\langle v(x,t)\rangle$ decay exponentially as $\sim \gamma^t$, corresponding to an effective dephasing-induced decay rate $\kappa = -\ln \gamma$. Fig. 1e shows excellent agreement between the theoretical curve (solid line) and values extracted from simulated mean-field decays (circles). Insets display two representative field patterns: at smaller $W$ (larger $\gamma$), the field decays slowly, whereas at larger $W$ (smaller $\gamma$) it decays rapidly, visually confirming that $\gamma$ directly measures the system's degree of coherence. The factor $\Phi_{\alpha\beta}(x,n) = \exp[\delta_{\alpha u}g(x+n)+\delta_{\beta u}g(x-n)]$ encodes the site-dependent non-Hermitian potential $g(x) = \lambda\cos[2\pi\alpha(x+1)]$. The corresponding expression for a Hermitian (real) quasiperiodic potential is provided in Supplementary Note 1 and employed in our experiments and theoretical analysis of Hermitian quasicrystals (Figs. 2 and 3). The integer displacement functions $\Delta n(\alpha, \beta)$ and $\Delta x(\alpha, \beta)$ implement the appropriate index shifts for each tensor component, as dictated by the underlying lattice geometry.

Equation (2) constitutes a closed set of deterministic recurrence relations that describe the exact ensemble average over all random phase realizations. In the fully coherent limit ($W = 0$, $\gamma = 1$), it reduces to the unitary evolution of pure states. In the fully incoherent limit ($W = 2\pi$, $\gamma = 0$), all higher-order correlators ($n > 0$) vanish, and the dynamics collapse to a classical random walk governed solely by the densities, that is

$$\begin{aligned}\langle|u(x,t+1)|^2\rangle &= \cos^2(\theta)\langle|u(x+1,t)|^2\rangle + \sin^2(\theta)\langle|v(x-1,t)|^2\rangle \\ \langle|v(x,t+1)|^2\rangle &= \sin^2(\theta)\langle|u(x-1,t)|^2\rangle + \cos^2(\theta)\langle|v(x+1,t)|^2\rangle.\end{aligned} \tag{3}$$

For intermediate dephasing strengths ($0 < W < 2\pi$), our framework smoothly interpolates between the fully coherent and fully incoherent limits, providing a unified description of the system dynamics across the entire coherence landscape. This theoretical formulation not only captures the ensemble-averaged dynamics for both Hermitian and non-Hermitian settings, but also reveals how each physical ingredient, including coherent coupling, non-Hermitian potentials, and dephasing, collectively determines the system's transport properties. Remarkably, the theory shows excellent agreement with numerical simulations and accurately reproduces the dynamics across all dephasing strengths (see Extended Data Fig. 2), establishing a unified description of propagation across the entire Hermitian-coherence landscape.

**Non-diffusive transport in fully decoherent non-Hermitian quasicrystals**

Next, we experimentally investigate how wave transport in disordered lattices is governed by the interplay of coherence and non-Hermiticity. We start our experiments by probing the delocalization-localization phase transition in a Hermitian quasicrystal (the Aubry–André model) under coherent conditions ($W = 0$). To characterize the localization property of the system, we excite the system at a single site, and measure the inverse participation ratio (IPR) of the wave-packet profile after $t = 80$ time-steps. The IPR is defined as[35,41] $\mathrm{IPR} = \sum_x[P(x,t)]^2$, where $P(x,t) = [|u(x,t)|^2+|v(x,t)|^2]/\sum_x(|u(x,t)|^2+|v(x,t)|^2)$ is the normalized intensity distribution. Physically, IPR represents the inverse of the number of sites occupied by the eigenstate. For extended states, IPR is small and vanishes as $\sim 1/L$ with system size $L$, whereas for a tightly-confined eigenstate, the IPR is independent of $L$ and takes a finite value close to 1 (see Supplementary Note 3). The conventional IPR is typically applied to eigenstates, here we evaluate the same quantity on the time-evolved wave packet. As the coupling ratio $\theta$ increases, a sharp delocalization-localization transition is observed (Fig. 2a). To further characterize the full profile of this phase transition, we plot the IPR curve measured at a much longer evolution time ($t = 2000$). The curve clearly shows a sharp jump in IPR near a critical point $\theta_c \simeq 0.45\pi$, reinforcing the abrupt transition from extended to localized regimes. The corresponding field patterns clearly distinguish between extended from localized dynamics. To classify the transport type in each phase, we analyze the wave-packet spreading via its second moment $M_2(t) = \sum_x x^2 P(x,t)$. In the long-time limit, $M_2(t)$ follows a power-law growth $M_2(t) \sim t^k$. The exponent $k$, or the spreading coefficient, serves as a dynamical indicator: $k = 2$ corresponds to ballistic transport, $k = 1$ to diffusive transport, and $k = 0$ to localization. As shown in the upper panel of Fig. 2a, dynamically, the phase transition in a coherent Hermitian quasicrystal corresponds to a change from ballistic spreading to localization. This result is consistent with earlier reported experimental demonstrations of the phase transition in the Hermitian optical Aubry–André model[42]. In contrast, when strong dephasing ($W = 2\pi$) is introduced, the transition is completely washed out: the dynamical IPR curve flattens, and the wave packet exhibits diffusive spreading ($k = 1$) regardless of $\theta$ (Fig. 2b). The insets display two representative field patterns; although the diffusion rate varies with $\theta$, the transport remains diffusive. This observation confirms the conventional understanding that dephasing destroys Anderson localization in Hermitian systems[18].

Then, we study the wave transport in non-Hermitian quasicrystals. Under coherent conditions ($W = 0$), the non-Hermitian quasicrystal also exhibits a clear delocalization-localization transition (Fig. 2c). The transition occurs at a critical point $\theta_c$ that is nearly identical to that of the Hermitian case, which can be explained by a self-duality relation enabled by the imaginary quasiperiodic potential (see Supplementary Note 2). The transport behavior in the extended regime ($\theta < \theta_c$) differs fundamentally between the Hermitian and non-Hermitian cases. In the coherent Hermitian quasicrystal, the transport is ballistic ($k = 2$). In contrast, the non-Hermitian quasicrystal exhibits a two-step dynamical evolution. At intermediate times, the system first undergoes a transition from ballistic ($k = 2$) to diffusive ($k = 1$) spreading, a hallmark of non-Hermitian transport that arises from mode-competition and spectral narrowing, analogous to the dissipation-induced mobility transition reported in ordered non-Hermitian lattices[43]. The transition is clearly evidenced by the distinct short-time (blue dashed line in Fig. 2c, upper panel) and long-time (solid line) spreading coefficients that approach $k = 2$ and $k = 1$, respectively (see Extended Data Fig. 3 for the full evolution of $M_2(t)$).

Surprisingly, when full decoherence ($W = 2\pi$) is introduced, a transport transition remains clearly visible (Fig. 2d), in sharp contrast to the featureless diffusive transport observed in decoherent Hermitian systems. Specifically, as the coupling strength $\theta$ is varied, the system undergoes a sharp transition from diffusion to localization despite the complete absence of phase coherence. Moreover, although both coherent and incoherent non-Hermitian systems may exhibit diffusive transport on the delocalized side of the transition, the physical origin of diffusion in these two regimes is fundamentally different. In the coherent case, the dynamics exhibit a characteristic two-step evolution: diffusion emerges dynamically from an initial ballistic propagation and retains finite off-site correlations associated with residual phase coherence. By contrast, in the fully incoherent regime transport is purely diffusive from the outset, and the dynamics reduce to a memoryless Markov process governed solely by on-site densities. This distinction manifests spatially in the interference-rich profiles characteristic of coherent diffusion versus the featureless Gaussian profiles observed under strong dephasing, and temporally in the immediate onset of steady diffusion in the incoherent case. Taken together, these observations establish a distinct transport paradigm in which decoherence actively reshapes non-Hermitian dynamics by enabling a genuine diffusion-localization transition, rather than merely suppressing coherent interference.

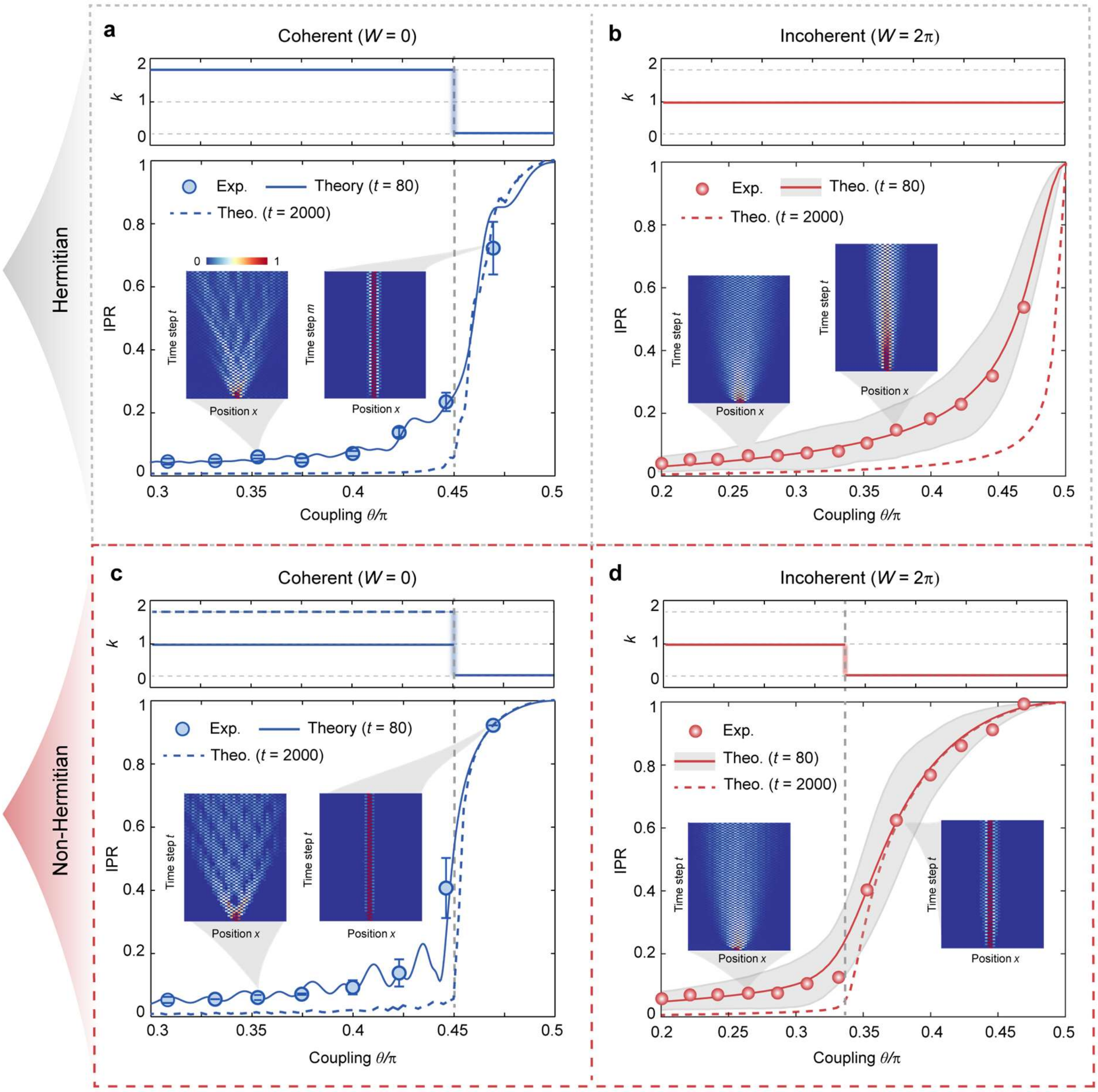

**Fig. 2. Diffusion-localization transition in fully-decoherent non-Hermitian quasicrystals.** (**a**) Hermitian quasicrystals under coherent conditions. Upper panel: long-time spreading coefficient $k$ as a function of coupling strength, characterizing the asymptotic transport behaviour. The shaded region indicates the critical regime near the transition point. lower panel: the inverse participation ratio of the evolved wave packet reveals a sharp localization transition. Insets: field evolution from a single-site excitation, showing extended (left) and localized (right) dynamics on either side of the transition. The dashed line represents the critical point of transition. (**b**) Same Hermitian system under strong dephasing. The transition is completely destroyed, with the IPR curve flattening and the system exhibiting diffusive transport regardless of dephasing strength (insets). Red circles are the experimental results averaged over 30 different dephasing configurations. For a large number of dephasing configurations, the IPR should converge to the red line, which is derived from Eq. (3). The grey shadow around the red line shows ±1 standard deviation of expected statistical

fluctuations, averaged on 400 realizations. (**c**) Non-Hermitian quasicrystals under coherent conditions. The blue dashed line in the upper panel corresponds to the short-time spreading coefficient. A localization transition occurs, evidenced by a jump in IPR. Insets illustrate the corresponding delocalized and localized wave dynamics. (**d**) Same non-Hermitian quasicrystals under dephasing. Contrary to the Hermitian case, the localization transition survives.

**Phase diagram of decoherent non-Hermitian quasicrystals**

To systematically explore the interplay between non-Hermiticity and decoherence, we employ our unified decoherent theory to construct comprehensive phase diagrams of the localization landscape as a function of both the coupling strength $\theta$ and the dephasing strength $W$. We first compute the eigenstate-based phase diagram that reflects the actual long-time dynamics of the system. Since the asymptotic behavior is dominated by the eigenstate with the largest eigenvalue (the longest-lived asymptotic state), we quantify its spatial localization using the asymptotic-state inverse participation ratio (AS-IPR), defined as the conventional IPR of this dominant eigenstate. The resulting phase diagram, shown in Fig. 3a, reveals a stark contrast between the two systems. For the Hermitian quasicrystal (left panel), the AS-IPR rapidly becomes uniformly low for $W > 0$, confirming that the asymptotic state is extended and no phase boundary persists under dephasing (dephasing destroys the spectral localization transition entirely). By contrast, the non-Hermitian quasicrystal (right panel) exhibits a robust phase boundary that persists up to the maximum dephasing strength $W = 2\pi$. Moreover, the transition boundary shifts continuously toward smaller $\theta$ as $W$ grows, meaning that stronger dephasing lowers the critical coupling required for localization (or equivalently, for a fixed $\theta$, localization can be achieved with a weaker quasiperiodic modulation $\lambda$). To connect this spectral picture directly to dynamical observations, we simulate the finite-time wave-packet evolution following a single-site excitation and extract the corresponding IPR after 200 time-steps (Fig. 3b). The resulting dynamical phase diagram closely mirrors the eigenstate-based one, confirming the spectral picture provides a reliable predictor to the dynamical localization landscape. For the non-Hermitian case, the eigenstate-based and dynamical phase diagrams are in excellent quantitative agreement, validating the self-consistency of our theoretical framework. For the Hermitian case, while the asymptotic state is extended, the simulated dynamical IPR can remain small within our finite evolution time ($t = 200$) because the spreading toward this uniform state becomes anomalously slow as $\theta$ increases.

To further characterize the dynamical transition, we examine horizontal cross-sections ($\theta$ as a parameter) of the phase diagram at fixed dephasing strengths. In the Hermitian system (Fig. 3c), under coherent conditions ($W = 0$), all eigenstates localize simultaneously beyond the critical point, giving a sharp step in the IPR profile, as visualized by the spatial profiles of three representative eigenstates. The associated phase profiles of localized eigenstates (grey circles) are relatively flat, consistent with the interference-derived phase patterns characteristic of Anderson localization. Under partially coherent conditions ($W = \pi$), the asymptotic eigenstate (with eigenvalue $\mu = 1$) is a uniform, extended profile, indicating the absence of a spectral localization transition. This non-decaying, maximally mixed state governs the long-time dynamics, ensuring that any Hermitian system under dephasing invariably evolves toward diffusion. In fact, any nonzero dephasing ($W > 0$) introduces such an extended state as the dominant mode, and the dephasing strength determines the time required to reach this diffusive fate (see Extended Data Fig. 3). At full dephasing ($W = 2\pi$), all eigenstates are extended, and the system behaves as a classical random walk which can be described by Eq. (3).

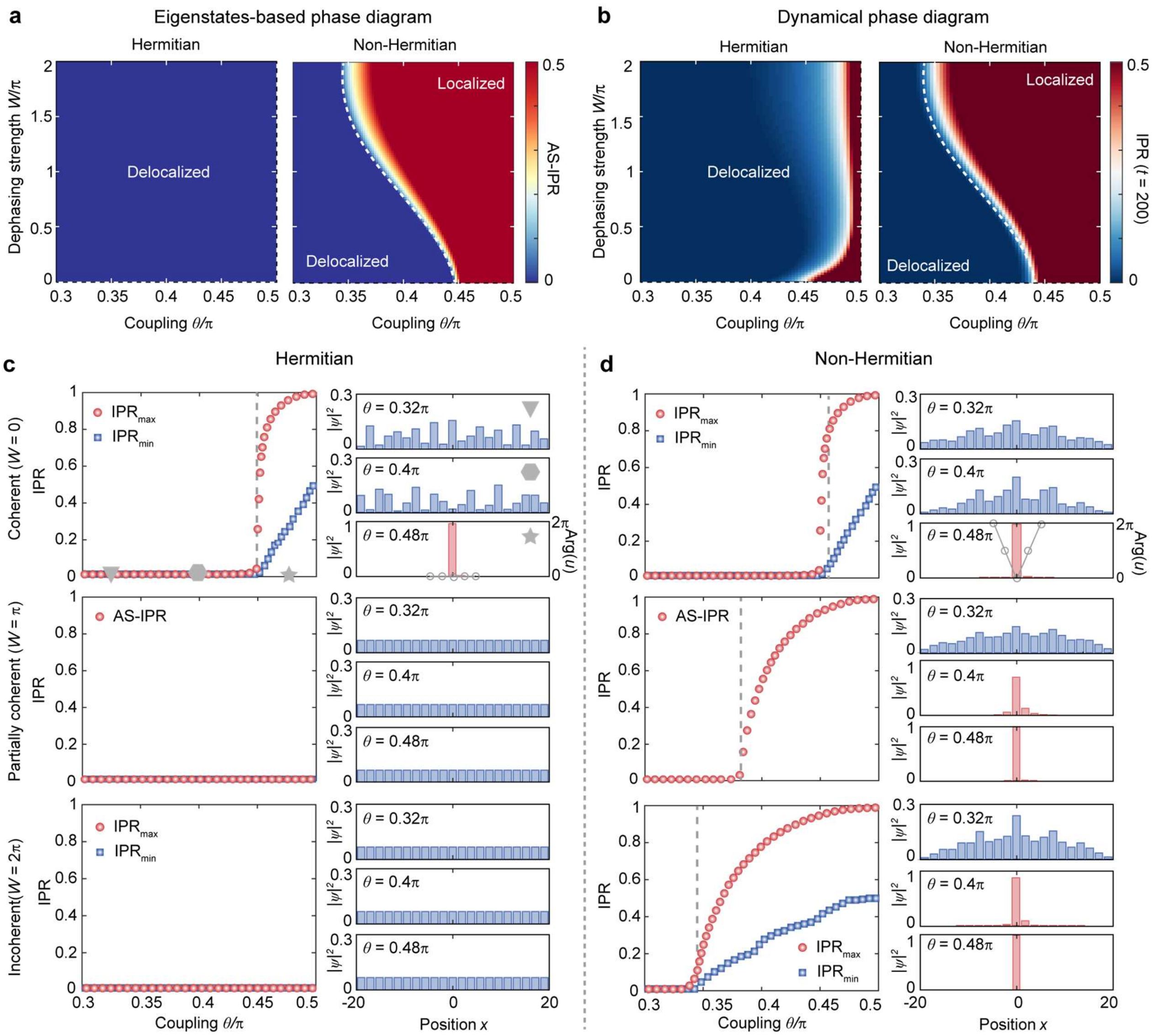

**Fig. 3. Phase diagrams of decoherent non-Hermitian quasicrystals.** (**a**) Eigenstate-based phase diagram in the ($\theta$, $W$) plane. Color represents the IPR of the longest-lived asymptotic state. Left: Hermitian quasicrystal. Right: Non-Hermitian quasicrystal with quasiperiodic dissipation. Two parameter lines are excluded from the IPR calculation and indicated by dashed lines: the fully coherent case $W = 0$ and the case of maximum coupling $\theta = \pi/2$. (**b**) Dynamical phase diagram from single-site excitation. Color maps the IPR of the field distribution after 2000 time-steps under the same ($\theta$, $W$) parameters as in (a). The strong agreement between (a) and (b) confirms that eigenstate localization predicts long-time dynamical behavior. (**c**) Hermitian quasicrystal. Cross-sections of the phase diagram at three dephasing strengths: coherent ($W = 0$, top), partially coherent ($W = \pi$, middle), and fully incoherent ($W = 2\pi$, bottom). Each row displays the IPR as a function of $\theta$ (left) and the spatial profile of the longest-lived eigenstate at three representative $\theta$ values: 0.32π, 0.4π, 0.48π (right), illustrating the transition (if exists) from extended to localized phases in the long-time limit. (**d**) Non-Hermitian quasicrystal. Corresponding cross-sections under the same dephasing strengths as in (c). Here the transition boundary shifts with increasing $W$, yet a clear step in IPR persists even under full dephasing. For the coherent cases, the phase profiles (grey circles) of the localized eigenstates are also shown for contrast.

For the non-Hermitian system (Fig. 3d), the steplike IPR profile, signaling a sharp localization transition, remains clearly visible even when dephasing is introduced. At $W = 0$, the transition resembles the Hermitian case where all eigenstates localize simultaneously. However, the underlying mechanism is qualitatively distinct: localization arises not from coherent interference but from a dynamic balance between diffraction and a non-reciprocal energy flow induced by the imaginary quasiperiodic potential. This key difference is directly evidenced by the phase profiles of the localized eigenstates (grey circles), which display organized gradients indicative of directional energy transport, in stark contrast to the interference-derived phase patterns of their Hermitian counterparts[31]. This distinction is crucial, as it implies the localization mechanism is inherently robust to phase randomization. Consequently, even as dephasing $W$ increases and global phase coherence is lost, the core non-Hermitian localization mechanism remains operative. At intermediate dephasing ($W = \pi$), the AS-IPR still shows a clear transition, but now the critical point is shifted. Ultimately, at full dephasing ($W = 2\pi$), all eigenstates still localize together beyond the shifted critical point, which stands in sharp contrast to the Hermitian case. This fundamental distinction between decoherent Hermitian and non-Hermitian systems is further elucidated by the tight-binding analysis in Supplementary Note 2, implying the generality of our conclusions across broader physical platforms.

**Decoherence-induced mobility edges and localization**

The continuous shift of the phase boundary in decoherent non-Hermitian quasicrystals points to a fundamental reorganization of the underlying eigenspectrum. Moreover, it suggests another intriguing possibility: dephasing can directly induce localization. Specifically, along a vertical trajectory in the phase diagram, where the coupling strength $\theta$ is held constant, increasing the dephasing strength $W$ drives the system from an extended phase into a localized regime. Such a mechanism, in which dephasing acts as a tuning parameter for localization, has no counterpart in conventional Hermitian disordered systems. Leveraging the unified theoretical framework established earlier, we now systematically trace how the spectral and dynamical localization properties evolve as dephasing is increased. To visualize this process, we fix the coupling at $\theta = 0.4\pi$ and track the eigenstate-resolved IPR across the spectrum, for increasing dephasing strength $W$ (Fig. 4a). Under fully coherent conditions ($W = 0$), all eigenstates are extended. As $W$ rises, a mobility edge emerges, separating the spectrum into extended and localized sectors. This mobility edge sweeps with increasing $W$, progressively converting extended eigenstates into localized ones. At full dephasing ($W = 2\pi$), all eigenstates become localized, resulting in a complete decoherence-induced spectral localization without any change in the static potential. This is striking because dephasing, traditionally viewed as a mere perturbation that suppresses localization, now acts as a continuous control parameter that deterministically sweeps a mobility edge and reshapes the entire localization landscape. We then directly measure the wave-packet dynamics at the two extremes of this transition. Following a single-site excitation, we measure the normalized intensity distribution across the lattice. Without dephasing ($W = 0$), the wave packet exhibits strong spatial spreading (Fig. 4b), in excellent agreement with the theoretical evolution from Eq. (2). The corresponding second moment $M_2(t)$ reveals a distinct two-step dynamics: the wave packet initially propagates ballistically ($k \approx 2$) before crossing over to a diffusive regime ($k \approx 1$) at longer times[43]. In stark contrast, under strong dephasing ($W = 2\pi$), the wave packet remains tightly confined near the excitation site (Fig. 4c), and $M_2(t)$ saturates rapidly to a low, constant value. The experimental data (dots) agree closely with the theoretical prediction from Eq. (3) (solid line), confirming that the average dynamics are accurately captured by the theory.

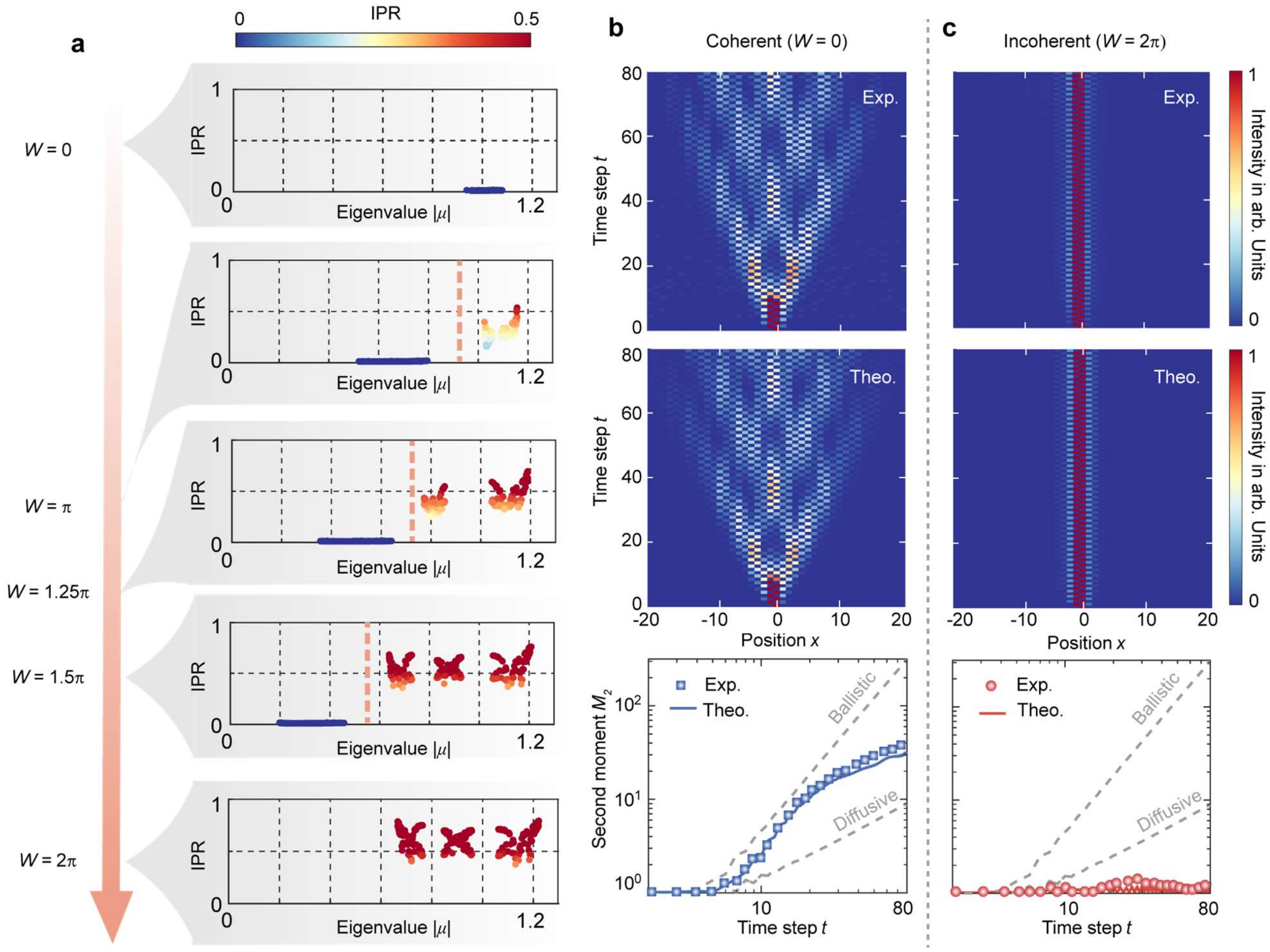

**Fig. 4. Decoherence-induced mobility edges and localization.** (**a**) Eigenspectrum evolution with increasing dephasing strength $W$. At $W = 0$ (top), all states are extended (low IPR). As $W$ increases, a mobility edge emerges and sweeps through the spectrum, progressively localizing eigenstates from the periphery inward. At $W = 2\pi$ (bottom), all eigenstates become localized, demonstrating complete decoherence-induced spectral localization. (**b**) Transport without dephasing ($W = 0$). Top: measured pulse propagation in the non-Hermitian quasicrystals. Middle: corresponding theoretical evolution from Eq. (2). Bottom: corresponding time evolution of the second moment $M_2(t)$. In the non-Hermitian quasicrystals, the wave dynamics exhibit a strong spatial spreading. The dashed lines, representing $M_2$ for ballistic and diffusive spreading, are calculated for the coupling ratio $\theta = 0.4\pi$ and shown for comparison. The slope of the curves in the logarithmic scale indicates whether the transport is ballistic or diffusive (the former is twice as much as the latter). (**c**) Transport under full dephasing ($W = 2\pi$). Top: measured pulse propagation in the non-Hermitian quasicrystals. Middle: theoretical prediction from the incoherent-walk model, Eq. (3). Bottom: corresponding time evolution of the second moment $M_2(t)$. The uncorrelated dephasing leads to strong localization. For a large number of realizations of random dynamic phase variations, the $M_2$ converges to the red line, which is calculated by Eq. (3).

## Conclusion

In conclusion, our work reveals that the long-held intuition linking decoherence to the restoration of diffusion is not universal, but rather a special property of Hermitian wave systems. By entering the decoherent non-Hermitian regime, we uncover a fundamentally different transport landscape in which the loss of phase coherence does not eliminate structure, but instead reshapes it. In non-Hermitian quasicrystals, decoherence does not merely erase interference; it reorganizes spectral lifetimes and dynamical pathways, enabling both decoherence-induced mobility edges at finite dephasing and diffusion-localization transitions in the fully incoherent limit. This behavior exposes a deeper physical principle: in open systems where transport is governed not only by interference but also by dissipation and mode competition, coherence is no longer the sole factor. Instead, decoherence emerges as an independent dimension of control, capable of driving phase transitions and stabilizing localized phases that are inaccessible in Hermitian settings. More broadly, our results establish decoherent non-Hermitian lattices as a distinct class of transport systems, governed by principles that lie beyond the Anderson paradigm. This perspective suggests that decoherence, traditionally regarded as an obstacle, can be deliberately harnessed as a resource when combined with non-Hermitian effects. Extending this approach to other non-Hermitian phenomena, such as the skin effect[35] or exceptional-point physics[44], may open access to entirely new regimes of wave control in realistic, noisy environments. As such, our findings provide a conceptual foundation for understanding and engineering wave dynamics in a wide range of open systems where dissipation and decoherence are inseparable.

## Data availability

All experimental data that were used to produce the results reported in this paper are available in the Zenodo repository at https://doi.org/10.5281/zenodo.18496302 (ref. 45).

## Code availability

The custom codes for this study that support the findings are available from the corresponding authors upon request.

## Methods

**Experimental setup.** The experimental setup for realizing the synthetic dimension, depicted in Fig. 1c, initiates each run by injecting a single 50 ns pulse into a longer fiber loop, which is connected to a shorter fiber loop through a VBS. The input pulse is cut out of a continuous wave signal from a tunable laser with an AOM with a 40 dB suppression ratio. Pulses traveling through the shorter loop are advanced in time, whereas those in the longer loop are delayed. This temporal shift, whether a delay or an advancement, can be thought of as equivalent to a spatial distance in the space dimension. At the same time, the number of round-trips introduces a temporal degree of freedom, which can be interpreted as a time dimension. A MZM in the longer loop modulates pulse amplitudes to simulate disordered dissipation. Injection and extraction of light in the loops are managed by OC, with photodetectors measuring the light intensity. Single-mode fibers are used to extend the propagation time for each loop to approximately 50 μs. By adding a fiber optic patch cable in one of the loops, a propagation time difference of 158 ns is created, distinguishing the long and short loops. A phase modulator in the longer loop provides phase disorder fluctuating in time, emulating the dephasing effects. An erbium-doped fiber amplifier in each loop compensates for propagation losses (for example, insertion losses and detection losses) and allows us to maintain a high optical signal-to-noise ratio. For optical gain clamping, high-power continuous wave laser signals at 1,535 nm are injected into the amplifiers via an OC. Bandpass filters suppress optical noise from the amplified spontaneous emission and remove the laser signal that was injected for the optical gain clamping. All fiber components are designed for operation at 1,550 nm wavelength and employ standard single-mode fibers. The polarization state of the propagating light is controlled via mechanical polarization controllers. The polarization needs to be aligned in front of polarization-sensitive components to obtain a high interference contrast. Arbitrary waveform generators generate all electrical signals that drive the modulators.

**Parameter control and data acquisition.** The coupling strength $\theta$ in the photonic lattice, corresponding to the splitting ratio of the VBS, is electrically tunable. The quasiperiodic modulation amplitude λ controls the strength of the non-Hermitian (imaginary) or Hermitian (real) quasiperiodic potential, implemented via the MZM or PM, respectively. The actual value of λ is determined by mapping the applied voltage to the corresponding transmission (for the MZM) or phase shift (for the PM) based on their calibrated response curves. The quasiperiodic parameter $\alpha = (\sqrt{5}-1)/2$ is the inverse golden ratio. In our experiments, we systematically varied the voltage applied to the VBS from 2.0 V to 4.4 V in steps of 0.2 V. This voltage range maps to the following set of discrete coupling angles $\theta$: 0.200π, 0.221π, 0.242π, 0.264π, 0.286π, 0.308π, 0.331π, 0.352π, 0.374π, 0.400π, 0.422π, 0.446π, 0.469π. For each parameter set (λ, $\theta$, $W$), the time-resolved optical intensity within the loops was measured. The light signal was converted to a voltage signal by a photodetector, subsequently amplified, and digitized by a high-speed oscilloscope. The raw voltage traces were then transferred to a computer for processing. Data processing involved mapping the temporally sampled signals onto a discrete two-dimensional grid representing the photonic lattice site $x$ and time step $t$, based on the calibrated temporal scales of the fiber loops. The resulting intensity distributions are normalized for visualization and further analysis. The data presented in Figs. 2 and 4 are obtained from such comprehensive parameter scans, with each dataset averaged over multiple realizations of the dephasing configurations where applicable.

**Spectral analysis.**

In the fully coherent limit ($W$ = 0), the eigenstates and eigenvalues are given by the unitary evolution in Eq. (1); in the fully incoherent limit ($W$ = 2π), they reduce to the classical Markov process described by Eq. (3). For intermediate dephasing strengths, the ensemble-averaged dynamics is governed by the recurrence relation for the two-point correlation tensor in Eq. (2). The eigenstates discussed here therefore characterize the ensemble-averaged dynamics, which approaches a classical Markov process in the incoherent limit. From this formalism we extract the effective non-Hermitian/Hermitian propagator $M$ whose eigenvalues $\mu_i$ determine the lifetime of eigenmodes. To quantify the contribution of each eigenstate to the long-time wave-packet profile, we introduce a dynamically weighted inverse participation ratio (DW-IPR) defined as

$$\text{DW-IPR}=\sum_i S_i \text{IPR}_i, \qquad S_i = \sum_{t=0}^{T} |\mu_i|^t,$$

where $\text{IPR}_i$ measures the spatial localization of the $i$-th eigenstate of $M$. In the limit of a very large evolution time $T \to \infty$, the sum is dominated by the eigenstate with the largest eigenvalue $|\mu_{\max}|$ (i.e., the longest-lived asymptotic state). Hence, the DW-IPR converges to the asymptotic-state IPR (AS-IPR): $\lim_{T\to\infty} \text{DW-IPR}(T) = \text{AS-IPR}$. The DW-IPR therefore reflects which eigenstates dominate the asymptotic dynamics—those with the largest eigenvalue $|\mu_{\max}|$. Eigenstate-based diagrams in the ($\theta$, $W$) plane are computed by diagonalizing $M$ for each parameter set and evaluating the DW-IPR or AS-IPR. For dynamical diagrams: simulating the full stochastic wave evolution over many dephasing configurations and calculating the late-time IPR of the averaged profile.

## References

1. Anderson, P. W. Absence of Diffusion in Certain Random Lattices. *Phys. Rev.* **109**, 1492–1505 (1958).
2. Drude, P. Zur Elektronentheorie der Metalle. *Annalen der Physik* **306**, 566–613 (1900).
3. Sommerfeld, A. Zur Elektronentheorie der Metalle auf Grund der Fermischen Statistik. *Zeitschrift für Physik* **47**, 1–32 (1928).
4. Schwartz, T., Bartal, G., Fishman, S. & Segev, M. Transport and Anderson localization in disordered two-dimensional photonic lattices. *Nature* **446**, 52–55 (2007).
5. Segev, M., Silberberg, Y. & Christodoulides, D. N. Anderson localization of light. *Nat. Photon.* **7**, 197–204 (2013).
6. Sheinfux, H. H. *et al.* Observation of Anderson localization in disordered nanophotonic structures. *Science* **356**, 953–956 (2017).
7. Stützer, S. *et al.* Photonic topological Anderson insulators. *Nature* **560**, 461–465 (2018).
8. Wang, P. *et al.* Localization and delocalization of light in photonic moiré lattices. *Nature* **577**, 42–46 (2020).
9. Hu, H., Strybulevych, A., Page, J. H., Skipetrov, S. E. & Van Tiggelen, B. A. Localization of ultrasound in a three-dimensional elastic network. *Nat. Phys.* **4**, 945–948 (2008).
10. Billy, J. *et al.* Direct observation of Anderson localization of matter waves in a controlled disorder. *Nature* **453**, 891–894 (2008).
11. Roati, G. *et al.* Anderson localization of a non-interacting Bose–Einstein condensate. *Nature* **453**, 895–898 (2008).
12. Kondov, S. S., McGehee, W. R., Zirbel, J. J. & DeMarco, B. Three-Dimensional Anderson Localization of Ultracold Matter. *Science* **334**, 66–68 (2011).
13. Wiersma, D. S. The physics and applications of random lasers. *Nat. Phys.* **4**, 359–367 (2008).
14. Madhukar, A. & Post, W. Exact Solution for the Diffusion of a Particle in a Medium with Site Diagonal and Off-Diagonal Dynamic Disorder. *Phys. Rev. Lett.* **40**, 70–70 (1978).
15. Gurvitz, S. A. Delocalization in the Anderson Model due to a Local Measurement. *Phys. Rev. Lett.* **85**, 812–815 (2000).
16. Brun, T. A., Carteret, H. A. & Ambainis, A. Quantum to Classical Transition for Random Walks. *Phys. Rev. Lett.* **91**, 130602 (2003).
17. Broome, M. A. *et al.* Discrete Single-Photon Quantum Walks with Tunable Decoherence. *Phys. Rev. Lett.* **104**, 153602 (2010).
18. Schreiber, A. *et al.* Decoherence and Disorder in Quantum Walks: From Ballistic Spread to Localization. *Phys. Rev. Lett.* **106**, 180403 (2011).
19. Levi, L., Krivolapov, Y., Fishman, S. & Segev, M. Hyper-transport of light and stochastic acceleration by evolving disorder. *Nat. Phys.* **8**, 912–917 (2012).
20. Gopalakrishnan, S., Islam, K. R. & Knap, M. Noise-Induced Subdiffusion in Strongly Localized Quantum Systems. *Phys. Rev. Lett.* **119**, 046601 (2017).
21. Biggerstaff, D. N. *et al.* Enhancing coherent transport in a photonic network using controllable

decoherence. *Nat. Commun.* **7**, 11282 (2016).
22. Dahan, R. *et al.* Imprinting the quantum statistics of photons on free electrons. *Science* **373**, eabj7128 (2021).
23. El-Ganainy, R. *et al.* Non-Hermitian physics and PT symmetry. *Nat. Phys.* **14**, 11–19 (2018).
24. Xu, G. *et al.* Diffusive topological transport in spatiotemporal thermal lattices. *Nat. Phys.* **18**, 450–456 (2022).
25. Longhi, S. Incoherent non-Hermitian skin effect in photonic quantum walks. *Light Sci. Appl.* **13**, 95 (2024).
26. Jin, P. *et al.* Temporal anti-parity–time symmetry in diffusive transport. *Nat. Phys.* 1–7 (2025) doi:10.1038/s41567-025-03129-8.
27. Yao, S. & Wang, Z. Edge States and Topological Invariants of Non-Hermitian Systems. *Phys. Rev. Lett.* **121**, 086803 (2018).
28. Yokomizo, K. & Murakami, S. Non-Bloch Band Theory of Non-Hermitian Systems. *Phys. Rev. Lett.* **123**, 066404 (2019).
29. Zhang, K., Yang, Z. & Fang, C. Correspondence between Winding Numbers and Skin Modes in Non-Hermitian Systems. *Phys. Rev. Lett.* **125**, 126402 (2020).
30. Zhao, H. *et al.* Non-Hermitian topological light steering. *Science* **365**, 1163–1166 (2019).
31. Basiri, A., Bromberg, Y., Yamilov, A., Cao, H. & Kottos, T. Light localization induced by a random imaginary refractive index. *Phys. Rev. A* **90**, 043815 (2014).
32. Weidemann, S., Kremer, M., Longhi, S. & Szameit, A. Coexistence of dynamical delocalization and spectral localization through stochastic dissipation. *Nat. Photon.* **15**, 576–581 (2021).
33. Li, B., Chen, C. & Wang, Z. Universal non-Hermitian transport in disordered systems. Preprint at https://doi.org/10.48550/arXiv.2411.19905 (2024).
34. Regensburger, A. *et al.* Parity–time synthetic photonic lattices. *Nature* **488**, 167–171 (2012).
35. Weidemann, S. *et al.* Topological funneling of light. *Science* **368**, 311–314 (2020).
36. Marques Muniz, A. L. *et al.* Observation of photon-photon thermodynamic processes under negative optical temperature conditions. *Science* **379**, 1019–1023 (2023).
37. Yu, L. *et al.* Dirac mass induced by optical gain and loss. *Nature* **632**, 63–68 (2024).
38. Yu, D. *et al.* Comprehensive review on developments of synthetic dimensions. *Photonics Insights* **4**, R06 (2025).
39. Longhi, S. Robust Anderson transition in non-Hermitian photonic quasicrystals. *Opt. Lett.* **49**, 1373 (2024).
40. Longhi, S. Dephasing-Induced Mobility Edges in Quasicrystals. *Phys. Rev. Lett.* **132**, 236301 (2024).
41. Wang, S. *et al.* High-order dynamic localization and tunable temporal cloaking in ac-electric-field driven synthetic lattices. *Nat. Commun.* **13**, 7653 (2022).
42. Lahini, Y. *et al.* Observation of a Localization Transition in Quasiperiodic Photonic Lattices. *Phys. Rev. Lett.* **103**, 013901 (2009).
43. Eichelkraut, T. *et al.* Mobility transition from ballistic to diffusive transport in non-Hermitian lattices. *Nat. Commun.* **4**, 2533 (2013).

44. Bergholtz, E. J., Budich, J. C. & Kunst, F. K. Exceptional topology of non-Hermitian systems. *Rev. Mod. Phys.* **93**, 015005 (2021).
45. Ren, Y. *et al.* Dataset for 'Non-diffusion transport in decoherent non-Hermitian quasicrystals'. Zenodo https://doi.org/10.5281/zenodo.18496303 (2026).

**Acknowledgements**
This work was sponsored by the National Natural Science Foundation of China (NNSFC) under Grant U25D8017 (Y.Y.), Nos. 61975176 (H.C.), and 62401491 (H.X.); the Key Research and Development Program of the Ministry of Science and Technology under Grant Nos. 2022YFA1405200 (Y.Y.), 2022YFA1404900 (Y.Y.), 2022YFA1404704 (H.C.), 2022YFA1404902 (H.C.), and 2025YFA1412300 (H.X.); the Key Research and Development Program of Zhejiang Province under Grant No. 2022C01036 (H.C.); the Fundamental Research Funds for the Zhejiang Provincial Universities under Grant No. 226-2025-00231 (Y.Y.); and the Research Grants Council of the Hong Kong Special Administrative Region, China under Grant No. 24304825 (H.X.).

**Author contributions**
Y.Y., H.X., and Y.R. conceived the study and initiated the project. Y.R. designed the experimental setup and performed the experiments with assistance from R.Z., L.Z., and K.Y., Y.R. conducted the numerical simulations and data analysis. Y.R. and Y.Y. developed the theoretical framework. The manuscript was drafted by Y.R. and critically revised and edited by Y.Y. and H.X.. The research was supervised by Y.Y., H.C., and H.X.. All authors discussed the results, reviewed the manuscript, and approved the final version.

**Competing interests**
The authors declare no competing interests.

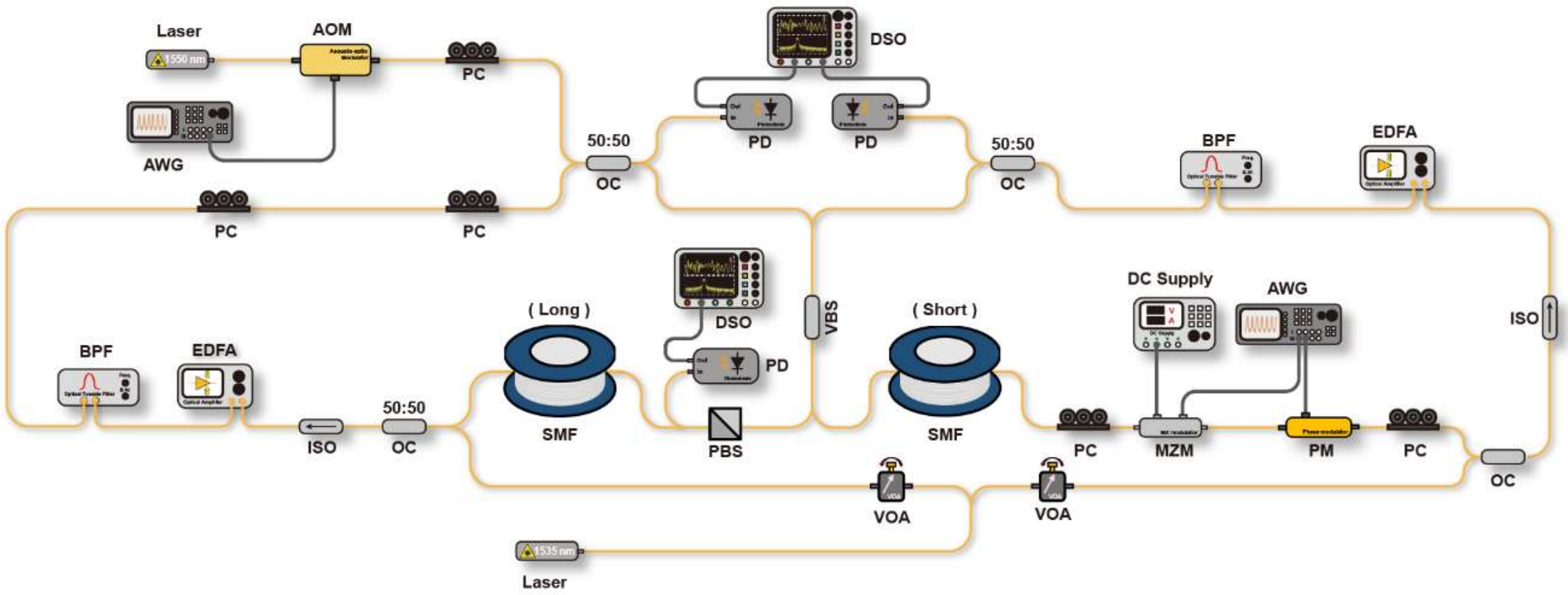

**Extended Data Fig. 1. Detailed experimental setup.** A pulse is created and injected into the left loop. Each fiber path comprises an erbium-doped fiber amplifier (EDFA) and 10 km of optical fiber. Mach-Zehnder modulators (MZM) and phase modulators (PM) allow for amplitude and phase modulation. AOM: acousto-optic modulator; BPF: bandpass filter; OC: optical coupler; VBS: variable beamsplitter; SMF: single-mode fiber; PD: photodetector; PBS: polarizing beam splitter; ISO: isolator; VBS: variable beamsplitter; AWG: arbitrary waveform generator; PC: polarization controller.

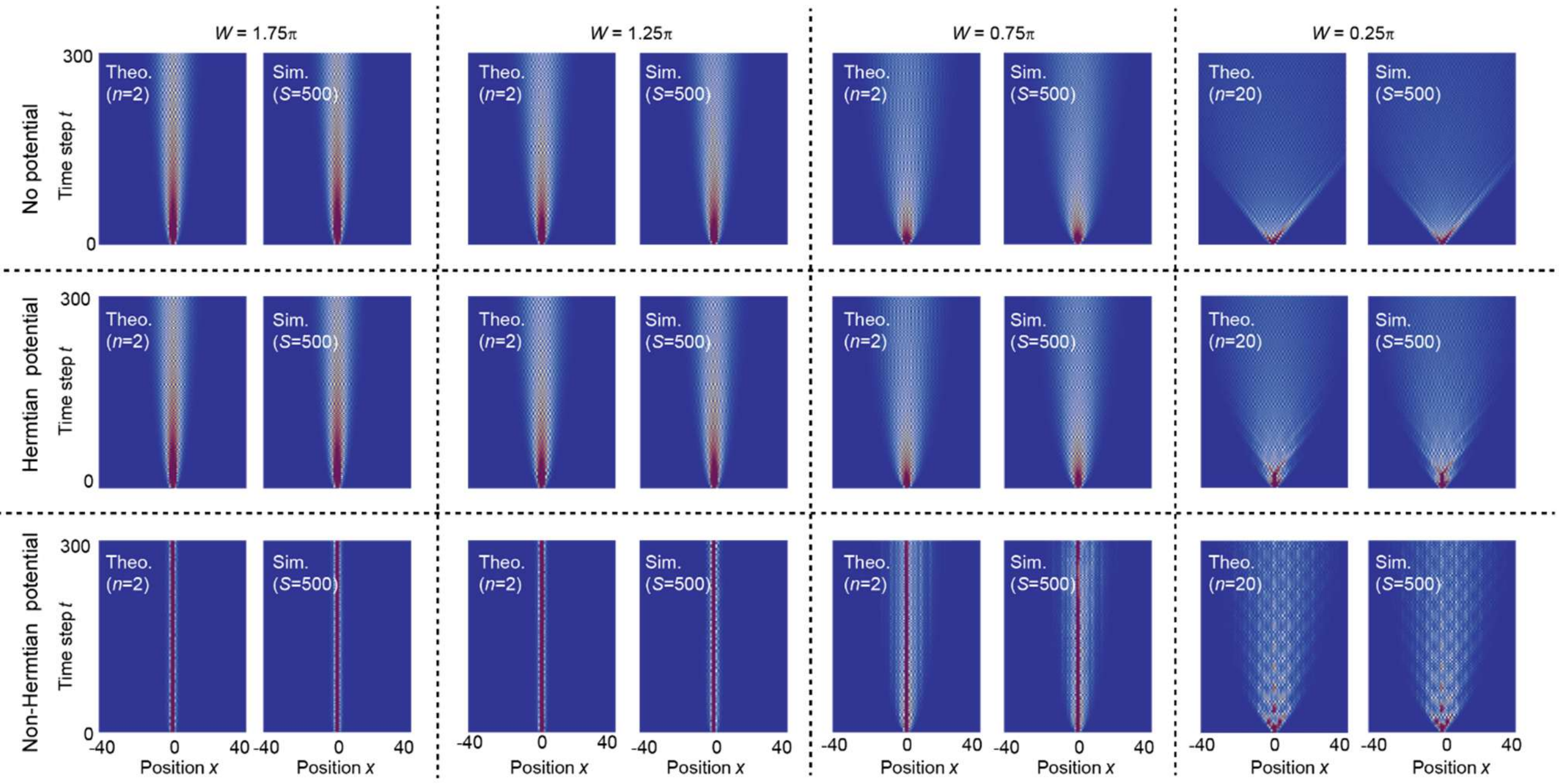

**Extended Data Fig. 2. Validation of the theoretical framework for decoherent systems.** Wave-packet dynamics are shown for a fixed coupling ratio $\theta$ = 0.8π under three on-site potential conditions: no potential, a Hermitian quasiperiodic potential, and a non-Hermitian quasiperiodic potential. For each case, results are presented for four dephasing strengths $W$ = 0.25π, 0.75π, 1.25π, 1.75π. Each panel compares two independent calculations: (left) direct evaluation using the theoretical recurrence relation Eq. (2), truncated at correlation order $n$ = 2; (right) Averaging of the full stochastic wave evolution over $S$ = 500 realizations using the original Eq. (2). Excellent agreement is observed across all parameter sets, confirming the accuracy of the unified decoherent theory. The quasiperiodic modulation strength λ is set to the same value used in the main-text experiments. Disorder is implemented by adding random dynamic phase shifts to the $u$ (short-loop) channel, while the excitation is applied in the $v$ (long-loop) channel, matching the experimental protocol.

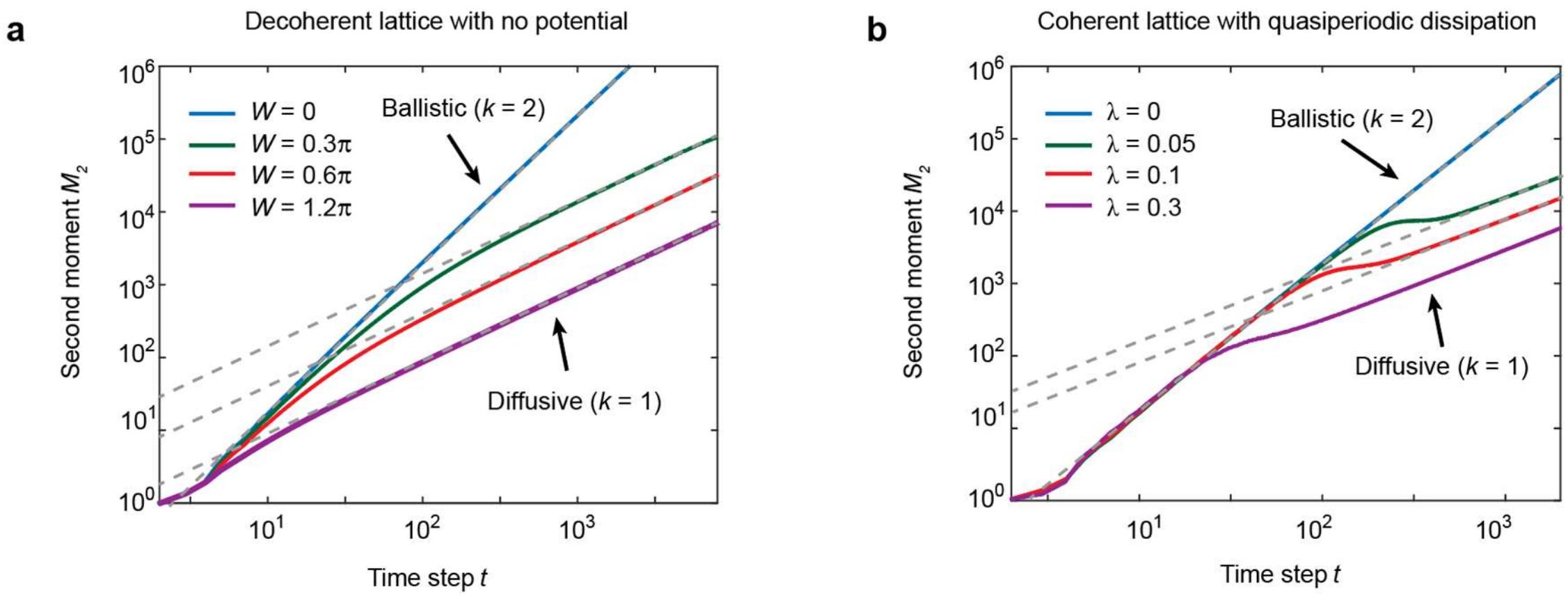

**Extended Data Fig. 3. Distinct pathways from ballistic to diffusive transport induced by dephasing and dissipation. (a)** $M_2(t)$ for a coherent lattice (no on-site potential) under increasing dephasing strength $W$. Without dephasing ($W = 0$), transport is ballistic ($M_2 \sim t^2$, dashed grey line). Introducing dephasing ($W > 0$) causes a gradual, monotonic crossover to normal diffusion ($M_2 \sim t$, dashed grey line). The crossover time decreases with increasing $W$, illustrating the continuous degradation of quantum coherence. **(b)** $M_2(t)$ for a lattice with quasiperiodic dissipation of strength $\lambda$ but no external dephasing. Even in the coherent limit, dissipation alone can induce a dynamical transition from ballistic to diffusive spreading. The transition occurs earlier as $\lambda$ increases, reflecting the mode-competition mechanism intrinsic to non-Hermitian dynamics. The dashed lines in both panels indicate slopes corresponding to ballistic ($k = 2$) and diffusive ($k = 1$) transport.

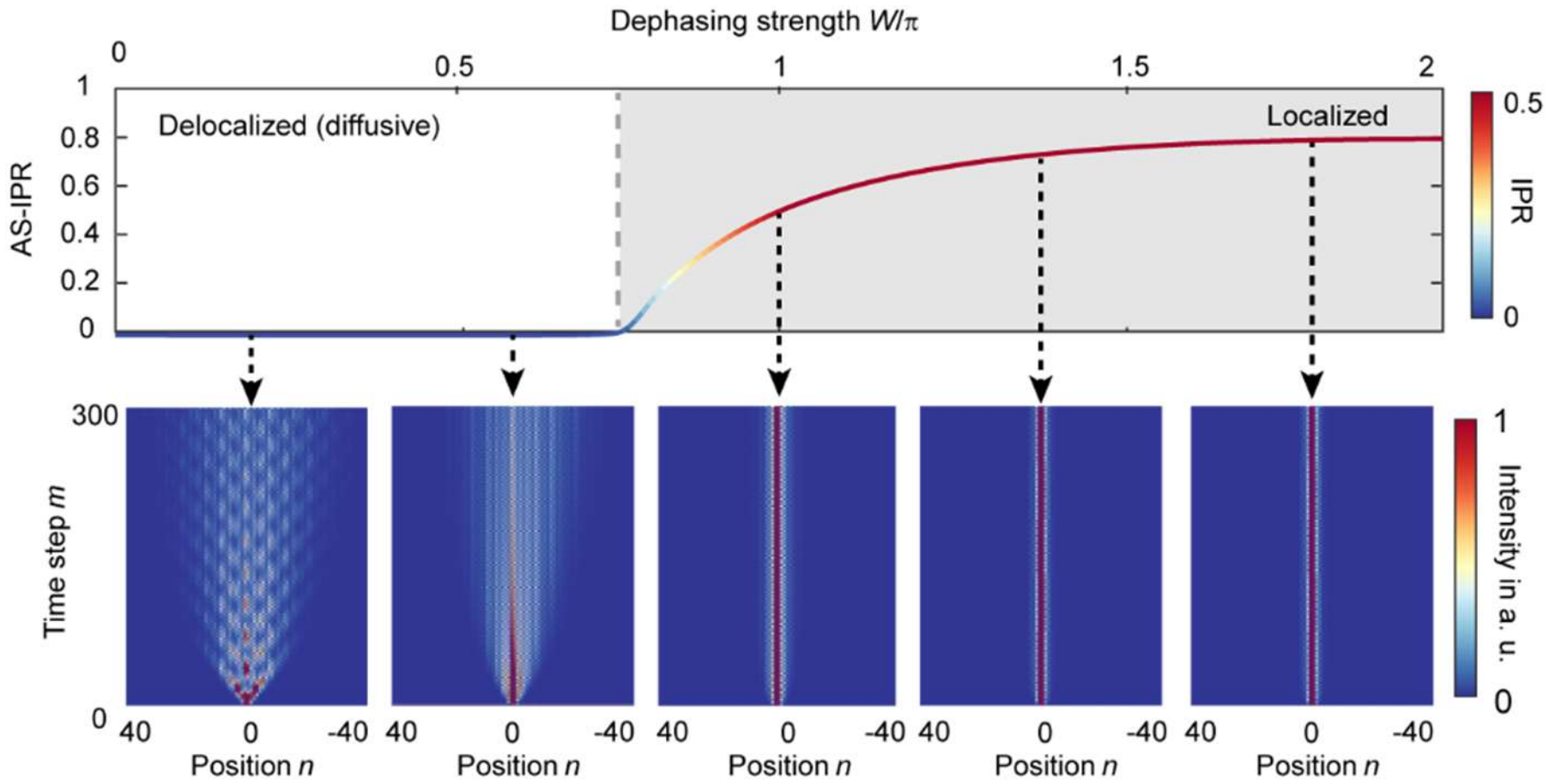

**Extended Data Fig. 4. Phase transition driven by dephasing strength.** Top: Asymptotic-state inverse participation ratio (AS-IPR) as a function of dephasing strength $W$ at fixed coupling $\theta = 0.4\pi$. The curve exhibits a sharp rise near $W \approx 0.75\pi$, marking a phase transition in the long-time limit. Bottom: Field profiles following single-site excitation at the five dephasing strengths. These panels provide a direct visual confirmation that the spectral measure (AS-IPR) accurately predicts the dynamical localization, and demonstrate that the dephasing strength $W$ acts as a continuous control parameter that sweeps the system from extended to fully localized dynamics.

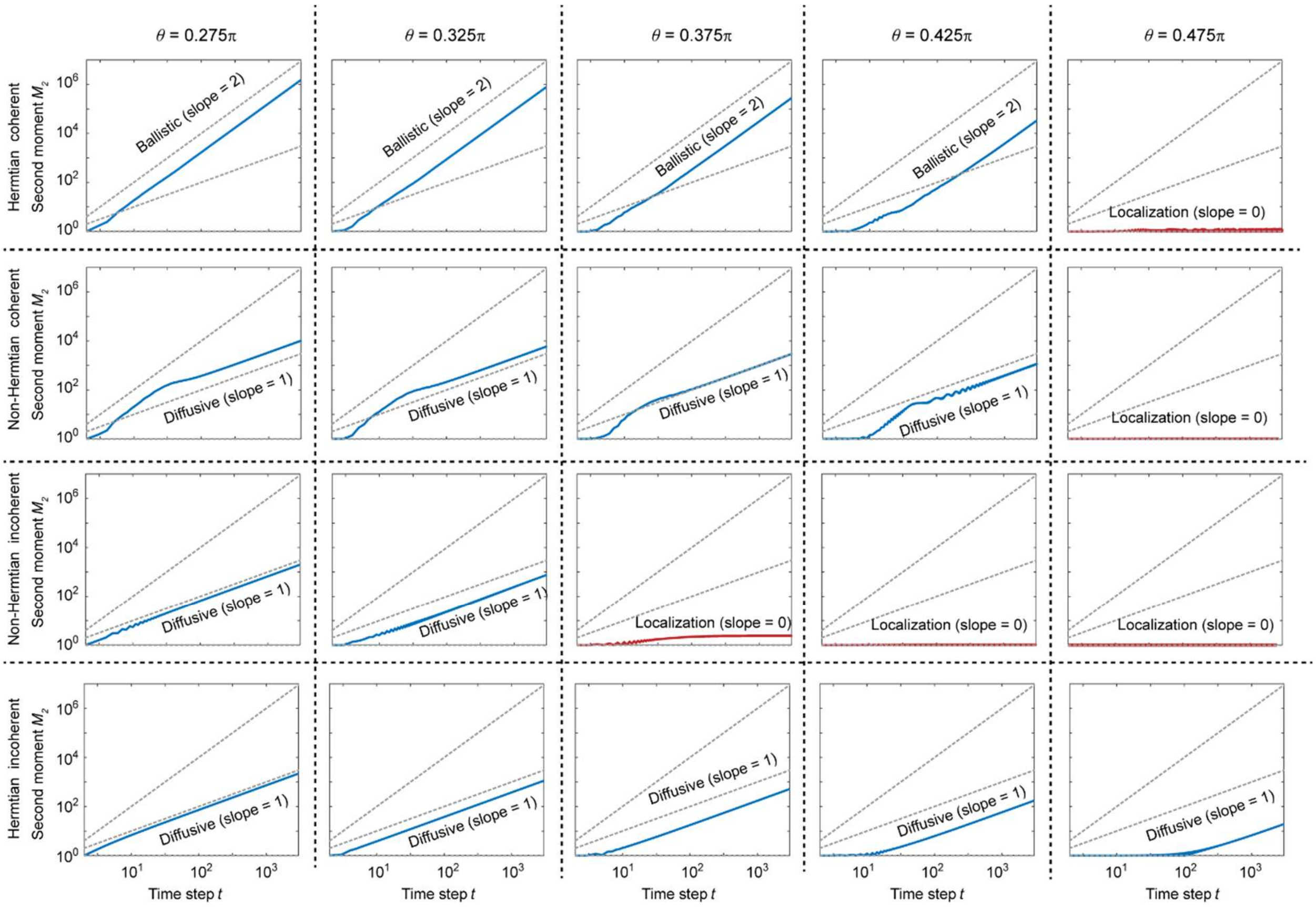

**Extended Data Fig. 5. Dynamical spreading across coherence and Hermiticity regimes.** Time evolution of the second moment $M_2(t)$ for five coupling strengths $\theta$ (columns) under four representative conditions (rows). Top row: coherent Hermitian lattice. Second row: coherent non-Hermitian lattice. Third row: fully incoherent non-Hermitian lattice ($W = 2\pi$). Bottom row: fully incoherent Hermitian lattice ($W = 2\pi$). Each panel is labelled with the observed asymptotic transport type (ballistic, diffusive, or localized) as determined from the long-time growth exponent $k$.